\documentstyle[twocolumn,aps,epsf]{revtex}
\newcommand{\ignore}[1]{}

\begin{document}
\twocolumn[
\hsize\textwidth\columnwidth\hsize\csname @twocolumnfalse\endcsname
\draft
\title{Phonon-assisted tunneling in an isolated double dot system } 

\author{J. Martorell$^a$, D.W.L. Sprung$^b$, GuoHong Yun$^{b,c}$ }
\address{Departament d'Estructura i Constituents de la Materia, Facultat
F\'{\i}sica, University of Barcelona\\  Barcelona 08028, Spain}
\address{$^b$Department of Physics and Astronomy, McMaster University, 
Hamilton ON L8S 4M1 Canada}
\address{$^{c}$Department of Physics, Inner Mongolia University, 
Hohhot, 010021, P.R. China}

\date{\today}
\maketitle

\begin{abstract}
Phonon-assisted tunneling rates are evaluated for a well isolated 
double dot system defined in a GaAs semiconductor heterostructure of 
finite thickness. A separable model for the confining potential 
allows accurate determinations of doublet electron wavefunctions and 
energies. It is found that at small doublet energies the 
piezoelectric rates due to flexural modes give the dominant 
contribution. For small slab thicknesses the predicted rates are up 
to two orders of magnitude larger than for very thick slabs. 
 \end{abstract}
\narrowtext
]

\section{ Introduction}

The problem we consider is the transfer of an electron between two 
dots separated by a tunneling barrier. The experimental situation we 
have is mind is that of two contiguous dots defined by electrostatic 
confinement, as in the experiments of references 
\cite{Fuj98,Qin02,Gar03,Hay03,DiC04}.  For open dots, where a current 
can be measured between leads connected to each dot, the problem has 
been well studied both experimentally and theoretically 
\cite{Bra99,Deb02,Bra04}.  More recently 
however, experimental results have been obtained for an isolated 
double dot system in a GaAs heterostructure \cite{Gar03}. 
Transfer was forced by varying the gate voltages on one of the 
dots, and charge transfer detected by a non-invasive voltage probe. 
Since this system has obvious interest as a possible experimental 
realization of a  q-bit and or as a half cell of a Quantum Cellular 
Automaton, a good understanding of all mechanisms leading to charge 
transfer is desirable. 

The simplest mechanism for transfer is direct tunneling, also called 
elastic tunneling: for a critical arrangement of the gate voltages 
some electron level has a wavefunction equally spread over the two 
dots, and slight changes in the applied voltage suffice to confine 
the wavefunction to one or the other of the two. For this to 
occur it appears necessary to have a doublet of electron states 
degenerate 
in energy at the critical configuration. However, initial 
experimental studies already showed the importance of inelastic 
tunneling,  mediated by acoustic phonons, when the energy separation 
between the doublet states is  approaching the minimum \cite{Fuj98}. 
Although the present generation of experiments has been carried out on two 
dimensional electron gases (2DEG) defined in very thick 
heterostructures, similar experiments may be done in future 
\cite{Smth} on 2DEG defined in heterostructures consisting of much 
thinner slabs. It is of interest to examine the effect 
that changes in the spectrum of phonon modes, due to finite slab 
thickness, would have on the phonon-assisted tunneling rates. A study 
along these lines has recently appeared for open double dots 
\cite{Deb02}. The purpose of the present work is complementary: we 
will address the problem of a strongly isolated  double dot system 
using a  more  realistic model for the confining potential. 

\section{Two coupled dots: separable model} 
\begin{figure}                  
\begin{center} 
\leavevmode 
\epsfxsize=8cm \epsffile{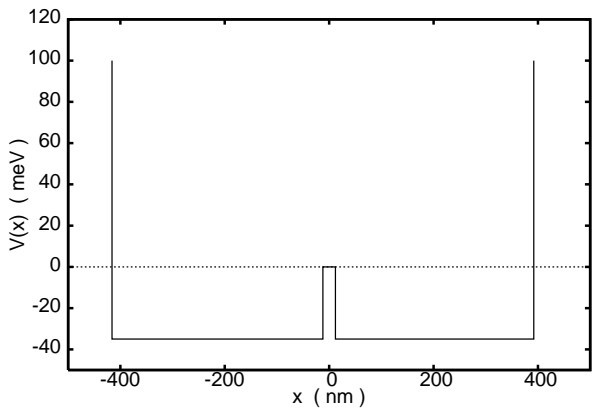} 
\end{center} 
\caption{ Confining potential profile along the $x$ direction.} 
 \label{fig1}
\end{figure}

In an earlier study\cite{Mar01}  on a device similar to that in the 
experiment of Gardelis {\it et al} \cite{Gar03}, we found that to 
good approximation a strongly isolated dot was well described by a 
confining potential of roughly rectangular shape and constant depth 
surrounded by very steep barriers. To account for leakage of 
electrons through one side of the dot it was sufficient to use a 
barrier of finite height and width derived from Poisson Thomas-Fermi 
simulations. Guided by the success of that model in explaining the 
lifetimes of the emitted electrons, we will use here a similar 
confining potential. For simplicity  and to achieve satisfactory 
numerical accuracy, we have chosen a rectangular shape for the 
barrier separating the two dots. Therefore, in the plane of the 2DEG 
we consider a double dot of rectangular shape, and define our $x$ and 
$y$ axes along the directions of its sides. We further set the 
confining potential, $V(x,y)$, to be infinite outside the boundaries of 
the rectangle and inside to depend only on $x$: the profile $V(x)$ 
is shown in Fig. \ref{fig1}. Dot $A$ on the left has sides $w_l 
\times l_y$, and dot $B$ on the right,  $w_r \times l_y$. The square 
barrier separating them runs from $-w_b/2$ to $w_b/2$. For 
convenience we set the zero of energy at the top of the barrier, so 
that inside dots $A$ and $B$ the potential is $V_D < 0$. The electron 
wavefunctions, in the envelope function approximation, 
$\Psi(x,y,z) = \phi(x) \psi(y) \chi(z)$, are obtained by solving a 3D 
separable Schr\"odinger equation.

 For simplicity, we chose the well known Fang-Howard \cite{Fang66} 
parametrization for $\chi(z)$, with parameters appropriate for the 
Cavendish experiment. We give its explicit form in Section III. 
Due to the simple form of the confining potential along $y$, the 
$\psi(y)$ are the well known eigenfunctions of the infinite square 
well. Along the $x$ axis: 
 \begin{equation}                       
-{\hbar^2\over {2m^*}} \phi''(x) + V(x) \phi(x) = E^{(x)}\phi(x)  
\label{eq:q1}
\end{equation}
and $V(x) = V_D $ when $ -w_l-w_b/2 < x < -w_b/2$ and  $ w_b/2 < x < 
w_b /2+ w_r $, whereas $V(x) = 0$ when  $ -w_b/2 < x < w_b/2$. 
The obvious analytic 
forms of $\phi(x)$ in the ``zones" (A, B or barrier) are matched at 
the boundaries (wavefunction and derivative) leading to 
explicit equations for the energy that are easily solved 
numerically to the required accuracy, even for sizeable barriers like 
those studied here. Once these eigenenergies have been determined it 
is a simple matter to find the fraction of the normalization of the 
wavefunctions in each dot. The assignment of each solution to a level 
in dot $A$ or $B$ is in most cases unambiguously determined 
by the fractions of normalization. By counting the number of nodes in 
the corresponding dot we assign to each solution an effective quantum 
number $n_l$ (left) or $n_r$ (right.)   The level crossings which 
occur for special values of the parameters are detected by the jumps, 
from almost zero to almost one, of those normalization fractions. 
There is always negligible normalization inside the barrier. 

The energy of a given level is the sum of the eigenvalues 
corresponding to the three eigenfunctions into which $\Psi(x,y,z)$ is 
factorized. However, in a doublet, the difference of energy 
between initial and final states has equal contributions from the $y$ 
and $z$ components since they must have the same $\psi(y)$ and 
$\chi(z)$.  Therefore in what follows $E_{if} = E^{(x)}_{Final}-
E^{(x)}_{initial}$. 

\begin{figure}                  
\begin{center}
\leavevmode
\epsfxsize=8cm
\epsffile{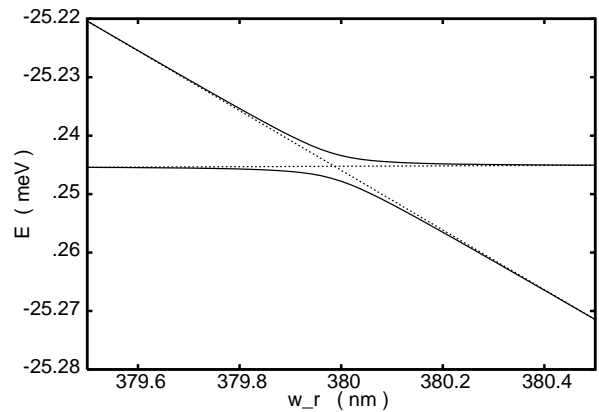}
\end{center}
\caption{ Energies. Doublet avoided level crossing, $n_r = 16$, $n_l 
= 17$.} 
 \label{fig2}
\end{figure}

 For definiteness, in the present  study we will use the width of  
dot B, $w_r$, as the parameter whose  variation simulates the changes 
induced by the variation of one of the confining gate voltages. In 
Fig. \ref{fig2} we show the level crossing chosen to present the 
results of our studies. The parameters  for the double dot 
configuration are: $V_D = -35 $ meV, $w_l = 404$ nm, $l_y = 350$ nm, 
$w_b = 24$ nm and $w_r$ is varied around $380$ nm. These parameter 
values are similar to those used in previous simulations of double 
dot devices also measured at the Cavendish \cite{Mar01,Quad}. 
The doublet shown corresponds to $n_l = 17$, $n_r = 16$ and with 
$\psi(y)= \sqrt{2/l_y} \cos (\pi y/ l_y)$, has a total 
energy close to the estimated Fermi level in the double dot system. 
The continuous lines in the figure are the doublet levels, whereas 
the dashed lines are inserted to guide the eye and would correspond 
to the ``unperturbed values'' of an ideal uncoupled double dot 
system. 

In a simple two-level system described by a hamiltonian: 
 \begin{equation}                       
H_{2l} = \pmatrix{ E_A & t \cr t & E_B} \ ,
\label{eq:q2}
\end{equation}
$\Delta E = \sqrt{(E_A-E_B)^2 + 4 t^2}$ is the separation of the two 
eigenstates. When the two unperturbed values cross, 
$E_A = E_B$, and $t = \Delta E/2$. This allows one to determine an 
effective tunneling width which characterizes the coupling between 
the two dots. From the data shown in Fig. \ref{fig2} we find $t = 
2.2\, \mu$eV.   

To simulate an  experiment 
we will assume for instance that the initial conditions correspond to 
$w_r = 380.4$ nm and that the lowest level in the right part of figure 
\ref{fig2} (corresponding to dot B) is occupied by one 
electron and the upper level in the same 
right part of the figure (dot A) is empty. We then 
reduce the value of $w_r$, shrinking the size of dot B but leaving 
dot A unchanged. The energy of the lowest level increases 
linearly as $w_r$ decreases so long as the occupation of the 
corresponding wavefunction stays close to unity in dot A, and stabilizes 
when  the electron tunnels to dot B at about $w_r = 380$ nm. This 
is direct or elastic tunneling. Inelastic tunneling due to phonon 
absorption will take place when $w_r$ is still above $380$ nm, via 
transitions from the lower level (electron in dot B) to the upper 
level (electron in dot A.) In our model we are not dealing explicitly 
with the coulomb energies of the electron in each well. For the 
configurations where the rates are maximal the initial and final 
electron wavefunctions are equally spread over the two dots, so 
we can assume that the contribution of Coulomb energy to the 
doublet separation cancels.   


\section{Acoustic phonon rates for a finite slab heterostructure} 
According to Fermi's golden rule, the rate for phonon
assisted tunneling is:
  \begin{equation}                      
w = {{2\pi}\over \hbar} |<\Phi_F \Psi_F | H_{e-ph} | 
\Phi \Psi>|^2 \ \delta(E-E_F \mp \hbar \omega) 
 \label{eq:sl1}
\end{equation}
where the bras and kets indicate initial and final states of the 
electron ($|\Psi>$) plus phonons ($|\Phi>$) and the $\mp$ sign 
applies to phonon emission or absorption.  Since GaAs is 
piezoelectric we will consider both the deformation potential and the 
piezoelectric contributions to the electron-phonon Hamiltonian. The 
specific forms of these rates for a finite slab have been thoroughly 
discussed in the work of Bannov {\it et al.}\cite{Bann94} which we 
follow closely. The hamiltonians describing both contributions can 
be written in second quantized form: 
 \begin{eqnarray}                       
&& H_{e-ph}^{(d,p)}  = \nonumber \\ && 
\sum_{{\vec q}_{\|}, n} \ e^{i{\vec q}_{\|}.{\vec 
r}_{\|}} \ \Gamma_{d/f/s}^{(d,p)}({\vec q}_{\|}, n, z) \left( 
c_n({\vec q}_{\|}) + c_n^{+} (-{\vec q}_{\|}) \right) 
 \label{eq:sl2}
\end{eqnarray}  
with the superscripts $d$ and $p$ indicating deformation and 
piezoelectric contributions and the subscripts $d$, $f$ and $s$ 
dilatational, flexural and shear modes. The z-axis is perpendicular to 
the plane of the 2DEG and the faces of the slab, and the origin is in 
the middle of the slab. Vectors in the $x,y$ plane are indicated by 
the subindex ${\|}$, and $n$ is the mode index. The expressions for 
the various $ \Gamma$'s are found in equations 33-34 (deformation 
potential) and 39-41 (piezoelectric) of \cite{Bann94}. 
Particularizing to the separable electron wavefunctions of our model: 
 \begin{eqnarray}                                 
w^{(d.p)} &=& {{2\pi}\over \hbar} \sum_{{\vec q}_{\|}, n} \ {\cal M}_{\|} \  \ 
|<\chi(z)| \Gamma_{d/f/s}^{(d,p)}({\vec q}_=, n, z)|\chi(z)>|^2 
\nonumber \\ && \times 
\left(N(\hbar \omega_n) + {1\over 2} \pm {1\over 2} \right) \ 
\delta(E-E' \mp \hbar \omega_n)\ , 
 \label{eq:sl3}
\end{eqnarray} 
where we have evaluated the matrix elements corresponding to the 
phonon occupations, and defined 
 \begin{equation}                                 
{\cal M}_{\|} \equiv |<\phi_F(x) \psi(y)|e^{i {\vec q}_{\|}.{\vec 
r}_{\|}} |\phi(x) \psi(y)>|^2~. 
 \label{eq:sl4}
\end{equation}
We remark that Bannov {\it et al.} \cite{Bann94} chose the axes in 
the 2DEG plane so that ${\vec q}_{\|} = ( q_x,0)$, whereas we have 
aligned the $x$ and $y$ directions along the sides of the double dot. 
Therefore, the $q_x$ appearing in their work correspond to $|{\vec 
q}_{\|}| \equiv q_{\|}$ in ours. Accordingly, we write:  
 \begin{eqnarray}                                  
{\cal M}_{\|} &=& |<\phi_F(x)|e^{i q_{\|} x \cos \theta_{\|}}|\phi(x)>|^2 
\nonumber \\  && \times \qquad 
|<\psi(y)|e^{i  q_{\|} y \sin \theta_{\|}} |\psi(y)>|^2 \ . 
 \label{eq:sl5}
\end{eqnarray} 
To show the coupling constants explicitly, we rewrite the $\Gamma$'s 
as follows: 
 \begin{eqnarray}                                 
\Gamma_{d/ f}^{(d)}({\vec q}_{\|}, n, z) &\equiv&  F_{d/ f, n}\ 
\sqrt{{\hbar E_a^2}\over {2A \rho \ \omega({\vec q}_{\|})}} \ 
\gamma^{(d)}_{d/ f, n}(q_{\|}, q_l, q_t; z) \nonumber \\ 
 \Gamma^{(p)}_{d/ f/ s} ({\vec q}_{\|}, n, z) &\equiv& F_{d/ f/ s,n}\  {{8 \pi 
e \beta}\over \epsilon} \sqrt{{\hbar \over 2 A \rho\ \omega_n({\vec 
q}_{\|})}} \nonumber \\ && \quad \times  \qquad
\gamma_{d/ f/ s, n}^{(p)}(q_{\|}, q_l, q_t; z) \ , 
 \label{eq:sl6}
\end{eqnarray}
with explicit expressions for the $\gamma$'s to be found in the 
above mentioned equations of \cite{Bann94}.  The normalization 
factors, $F_{ d/ f/ s}$ are given in their appendix. In 
this revised notation, the deformation rate becomes  
 \begin{eqnarray}                                 
w^{(d)} &=& \frac{1}{2 \pi\hbar}  \sum_{n; d,f} \int \ 
d{\vec q}_{\|}\  {\cal M}_{\|}\ F_n^2\ {{\hbar E_a^2}\over {2  \rho 
\omega_n({\vec q}_{\|})}} \nonumber \\ && \qquad \times  \quad 
|<\chi(z)|\gamma^{(d)}_n(z)|\chi(z)>|^2 
\nonumber \\ && \times 
\left(N(\hbar \omega_n) + {1\over 2} \mp {1\over 2} \right)\  
\delta(E-E_F \mp \hbar \omega_n) \ , 
 \label{eq:sl7}
\end{eqnarray}
where we have omitted the subindexes $d/f$ that appear each time a 
mode $n$ is referenced.  A similar expression applies to the 
piezoelectric rates. Furthermore,  for an infinite slab area, one has to  
replace $ 1/A \sum_{{\vec q}_{\|}} $ by the integral $(1/(2\pi)^2  
\int \ d{\vec q}_{\|} = (1/(2\pi)^2 \int \ q_{\|} \ d q_{\|} \ 
d\theta_{\|} $, so that 
 \begin{eqnarray}                                
&&\int_0^{2\pi}  d\theta_{\|}\  \int_0^{\infty} q_{\|} \ dq_{\|}\ 
\delta ( E- E_F - \hbar \omega ) \nonumber \\ && \qquad \qquad 
= \int_0^{2\pi} 
d\theta_{\|} \ q_{\|}  \left({{d \hbar \omega}\over {d 
q_{\|}}}\right)^{-1}~. 
\label{eq:sl8}
\end{eqnarray}
We then arrive at 
 \begin{eqnarray}                                
w^{(d)} &=& {\hbar \over {4\pi}} \sum_{n; d,f,s} {{E_a^2 \ q_{\|}}\over 
{\rho \ \hbar \omega_n \ ( d \hbar \omega_n/ dq_{\|})}}\ 
\left(N \hbar \omega_n + {1\over 2} \mp {1\over 2} \right) 
\nonumber \\ && \times \,\, 
F_n^2\, |<\chi(z)|\gamma^{(d)}_n(z)|\chi(z)>|^2 \  {\bar {\cal 
M}}_{\|} \ , 
 \label{eq:sl9}
\end{eqnarray}
where
 \begin{equation}                                
{\bar {\cal M}}_{\|} \equiv \int_0^{2\pi} \ d\theta_{\|} \ {\cal M}_{\|} \ .
\label{eq:sl10}
\end{equation}
To determine the expectation values of the $\gamma$'s we have chosen a 
Fang-Howard wavefunction to represent $\chi(z)$: 
 \begin{eqnarray}                                
\chi(z) &=&  {1\over \sqrt{2}} b^{3/2} ({\bar z} -{\bar z}_0) e^{-
b({\bar z} -{\bar z}_0)/2  } \quad , \quad {\bar z} > {\bar z_0} 
\nonumber \\ 
 &=& 0 \quad , \quad{\bar z} < {\bar z}_0 \ .
\label{eq:lg2}
\end{eqnarray}
with ${\bar z} = a/2- z$ and ${\bar z}_0$ is the depth of the 2DEG 
referred to the surface of the slab. We have chosen ${\bar z}_0 = 70$ 
nm and $b= 1/4$ nm$^{-1}.$ \\ 
We solved the algebraic equations 
(13),(14) and (18) of \cite{Bann94} numerically, so as to obtain the 
dispersion relation $\hbar \omega_n( q_{\|}) $ and corresponding 
expressions for the other momenta $q_t$ and $q_l$ appearing in the 
explicit expressions for the $\gamma^{(d,p)}_n(z)$ and the 
normalization factors $F_n^2$. 
\section{Results} 
\begin{figure}                   
\begin{center}
\leavevmode
\epsfxsize=8cm
\epsffile{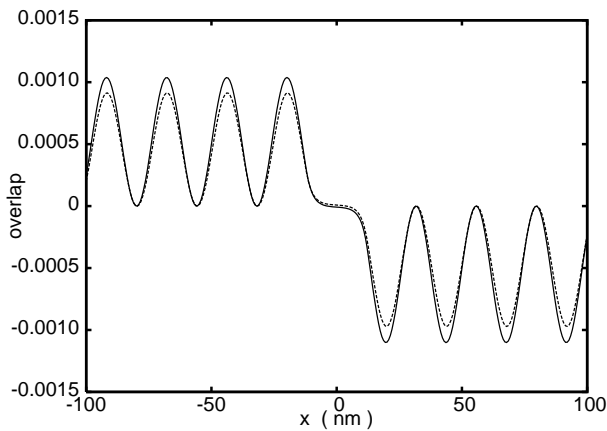}
\end{center}
\caption{ Overlaps of wavefunctions. Continuous line: the $\phi_F(x) 
\phi(x)$ of the  pair corresponding to $w_r = 379.8 $ nm. Dashed 
line: same for the pair at  $w_r = 380.2 $  nm. } 
 \label{fig3}
\end{figure}

Before presenting results for the rates, it is interesting to discuss 
some general features of the terms appearing in eq. \ref{eq:sl9} and 
their dependence on the doublet energy splitting, $E_{if}$. For a 
given mode, the corresponding dispersion relation $ E_{if} = \hbar 
\omega_n(q_{\|};a)$ determines the momenta $q_{\|}$, $q_t$ and $q_l$. 
As shown in figure \ref{fig2}, to a good approximation one finds the 
same value of the splitting energy for pairs of points located 
symmetrically with respect to the crossing point at $w_{r,cross} 
\simeq 380$ nm. One such pair is $w_{r,1} = 379.8 $ nm , $w_{r,2} = 
380.2 $ nm. For these two the values of $q_{\|}$ are nearly the same. 
This implies that all factors contributing to eq. \ref{eq:sl9} except  
${\bar {\cal M}}_{\|}$ will also be almost equal since they depend 
only on $q_{\|}$. The matrix element ${\bar {\cal M}}_{\|}$, involves 
the doublet wavefunctions $\phi(x)$ and $\phi_F(x)$ and these 
depend on the value of $w_r$.  It turns out however that due to the 
simplicity of the potential profile chosen for $V(x)$, the products of
wavefunctions for the paired doublets are also very similar. This is 
shown in Fig. \ref{fig3}, where we have plotted the overlaps
 $\phi_F(x) \phi(x)$ for the doublets $w_{r,1} = 379.8 $ nm ( continuous line )
and $w_{r,2} = 380.2 $ nm (dashed line). For clarity of the figure we only 
show the range $-100 < x < 100 $ nm, but the two lines practically coincide 
in most of the range of $x$'s. If we had chosen $w_{r,1}= 379.77 $ 
nm, so as to make the two $E_{if}$ even closer, the two lines in the 
plot would be indistinguishable 

The result of this symmetry is that the  
matrix element ${\bar {\cal M}}_{\|}$, is nearly identical for the 
two cases, since it is determined by the overlaps. In summary: the 
rates are very similar for pairs with $w_r$ symmetric with respect to 
the crossing point. Part of this is due to the simplicity of the 
potential profile chosen, but as long as the potential profile at the 
barrier is symmetric and each dot has a wide flat potential well, 
we expect that this symmetry will be preserved by less 
schematic double dot potential profiles.   \\ 
\begin{figure}                   
\begin{center}
\leavevmode
\epsfxsize=8cm
\epsffile{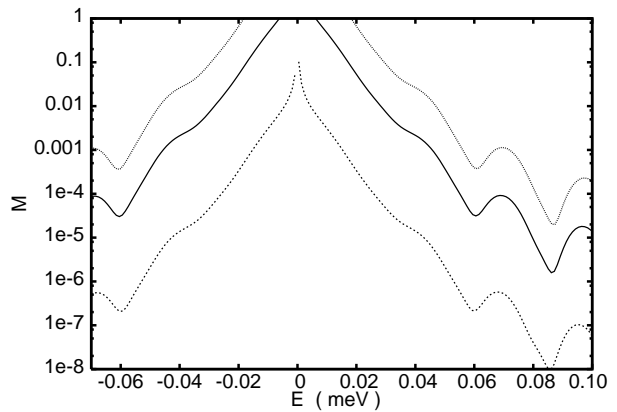}
\end{center}
\caption{ Matrix element ${\bar {\cal M}}_{\|}$ for various barrier 
widths: Continuous line $w_b = 24 $ nm ($t_{eff} = 2.2 \mu$eV) , 
dotted line: $w_b = 18 $ nm ($t_{eff} = 7.6 \mu$eV), dashed line: 
$w_b = 36$ nm ($t_{eff} = 0.2 \mu$eV).} 
 \label{fig4}
\end{figure}

\begin{figure}                   
\begin{center}
\leavevmode
\epsfxsize=8cm
\epsffile{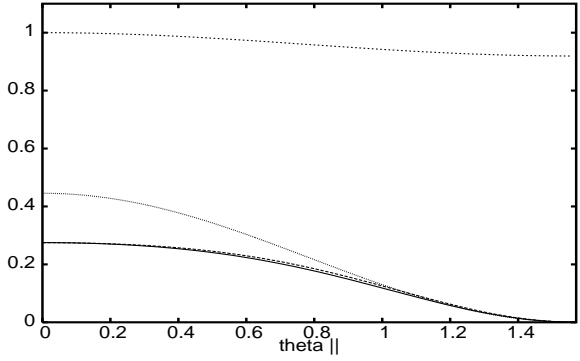}
\end{center}
\caption{The matrix elements entering ${\cal M}$ v.s. $\theta_{\|}$; 
two dot system with $w_r = 379.9 $ nm. 
Continuous line: Full ${\cal M}$. 
Close by dashed line: $|<\phi_F|e^{i q_{\|} \cos \theta_{\|} x}|\phi>|^2$. 
Short dashed line at top: $|<\psi_F|e^{i q_{\|} \sin \theta_{\|} y}|\psi>|^2$. 
Dotted line: approximate ${\cal M}$ of eq. {\protect \ref{eq:cm1}}. } 
 \label{fig5}
\end{figure}

Fig.\ref{fig4} shows ${\bar {\cal M}}_{\|}$ for three barrier widths. 
One sees that indeed the curves are symmetric with respect to $E_{if} 
=0$, and that the magnitude depends strongly on $w_b$. However, their 
shape depends little on the barrier width, showing the expected 
increase at small doublet energies. To analyze the latter, it is 
useful to consider the contributions to ${\bar {\cal M}}_{\|}$ for 
very low $q_{\|}$. When $q_{\|} \to 0$, eq. \ref{eq:sl5} can be 
approximated by 
 \begin{equation}                          
{\cal M}_{\|} \simeq q_{\|}^2 \cos^2 \theta_{\|} \ 
|<\phi_F(x)|x|\phi(x)>|^2  \ . 
\label{eq:cm1}
\end{equation}
Fig. \ref{fig5} shows these matrix elements ${\cal M}$ for a 
configuration with $w_r = 379.9 $ nm and $\psi(y)$ the lowest energy 
solution for the infinite square well. It can be seen that  $q_{\|} 
\simeq 4.5\ . \ 10^{-3}$ nm$^{-1}$ (corresponding to $E_{if} \simeq 
6.1 \mu$eV) is not sufficiently close to $0$ and therefore eq. 
\ref{eq:cm1} gives only a rough approximation for most values of 
$\theta_{\|}$. For other values of $w_r$ the agreement is similar. 
Eq. \ref{eq:cm1} is nevertheless useful because it correlates ${\cal 
M}_{\|}$ with the expectation value of $x$, and the behaviour of the 
latter is easier to visualize: 

For large doublet splitting the overlap of the wavefunctions 
$\phi(x)$ and $\phi_F(x)$ is small: one is mainly located in dot A, 
the other in B. As $E_{if}$ decreases each wavefunction begins to 
spread into the other dot. At the minimal doublet splitting, and 
assuming for sake of argument equal dot sizes, the 
wavefunctions become a symmetric and antisymmetric pair each with 
equal occupations in the two dots. The overlap is then maximal, but 
the product $\phi(x) \phi_F(x)$ is of positive sign in one dot and 
negative in the other so the resulting overlap integral vanishes. 
Introducing an $x$ in the integration, as in eq. \ref{eq:cm1}, makes 
the contributions from the two dots equal and therefore 
maximizes the matrix element when the two eigenfunctions are maximally 
spread over the two dots.

\begin{figure}                   
\begin{center}
\leavevmode
\epsfxsize=8cm
\epsffile{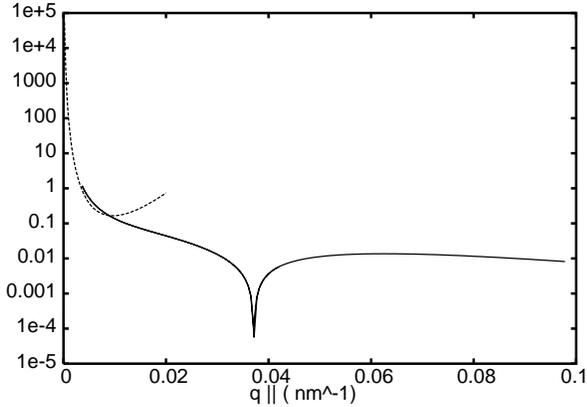}
\end{center}
\caption{ Matrix element ${\cal N}$ as a function of $q_{\|}$ for the 
lowest flexural mode. Continuous line: exact numerical calculation. 
Dashed line: small $q_{\|} $ expansion.} 
 \label{fig6}
\end{figure}

To complete the analysis, we show in Fig. \ref{fig6} the matrix 
element : $ {\cal N} \equiv | F_f <\chi(z)|\gamma^{(p)}_n(z)|\chi(z)>|$, 
for the lowest flexural mode. 
As will be shown in next section this mode gives the dominant 
contribution to the rate at small doublet splittings. At very small 
$q_{\|}$ one can perform power series expansions of the quantities 
entering into this matrix element and, neglecting the 2DEG thickness, 
one finds 
 \begin{equation}                           
{\cal N} \simeq {{18 s_l^2}\over {a^{9/2} s_t^2 }} 
{{\cosh(q_{\|} a/2) \cosh( q_{\|} (a/2-{\bar z}_0)) }\over 
( 1- (s_t/s_l)^2)^2\,\, q_{\|}^4}~, 
 \label{eq:gz1}
\end{equation}
where $s_l$ and $s_t$ are the velocities of bulk longitudinal and 
transverse waves. This approximation is shown as a dashed line in 
Fig. \ref{fig6} and reveals the cause of the sharp increase when 
$q_{\|} \to 0$. In conclusion, the two matrix elements, ${\bar {\cal 
M}}_{\|}$ and ${\cal N}$ contribute to the strong increase of the 
piezoelectric rates at small doublet energies, to be shown next. 

{\it Rates:} 
Following ref.\cite{Gar03}, we have chosen $T= 100 \ mK$ for all our 
rate calculations. Also, for definiteness, we have assumed that one of 
the symmetry axes of GaAs is parallel to $z$, whereas the other two 
are in the plane of the 2DEG, rotated by an angle $\theta_{sym} = 
\pi/5$ with respect to the $x,y$ axes introduced above. The material 
parameters and coupling constants for GaAs were taken from 
\cite{Glav02,Mitin99}. As discussed at the beginning of this section, 
to the accuracy of the logarithmic graphs to be presented, the 
rates for pairs of states with the same energy difference $E_{if}$  
are indistinguishable. To present emission and absorption 
rates in the same graph, we adopt the convention that negative 
(positive) $E_{if}$ correspond to phonon absorption (emission). 
Again, to the accuracy of the graphs, the difference between the two 
rates for the same $|E_{if}|$ is entirely due to the factor 
$N(E_{if}) + 1/2 \pm 1/2$.  

\begin{figure}                  
\begin{center}
\leavevmode
\epsfxsize=8cm
\epsffile{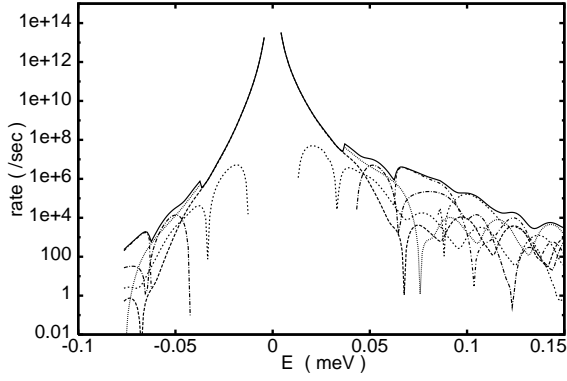}
\end{center}
\caption{ Piezoelectric rates due to flexural modes, as a function of 
doublet splitting. Continuous line sum of all 
contributions. Dashed lines :  separate contributions of each of the 
modes} 
 \label{fig7}
\end{figure}

\begin{figure}                  
\begin{center}
\leavevmode
\epsfxsize=8cm
\epsffile{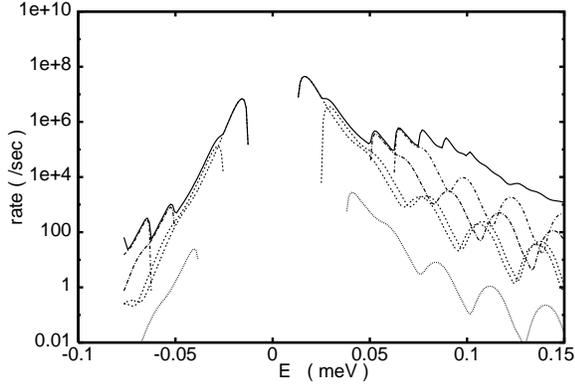}
\end{center}
\caption{ Piezoelectric rates due to shear modes, as a function of 
doublet splitting.} 
 \label{fig9}
\end{figure}

\begin{figure}                  
\begin{center}
\leavevmode
\epsfxsize=8cm
\epsffile{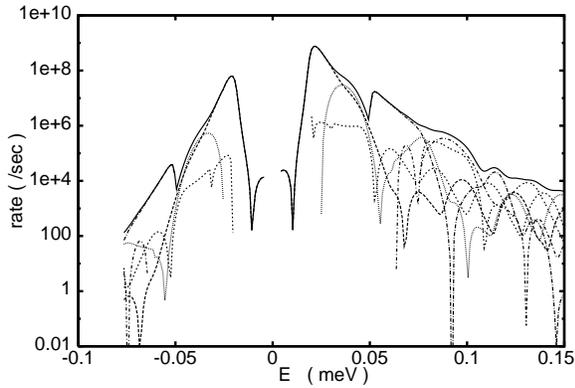}
\end{center}
\caption{ Piezoelectric rates due to dilatational modes, as a 
function of doublet splitting.} 
 \label{fig10}
\end{figure}

In Figs. \ref{fig7} to \ref{fig10} we present the rates for a slab 
of thickness $a = 500 $ nm, for the doublet shown in Fig. 
\ref{fig2}. The 2DEG is located at a depth ${\bar z}_0 = 70 $ nm  
below the surface and $\psi(y)$ is the lowest eigenfunction of the 
square well. Note the 
different scales in Fig. \ref{fig7} from the rest. This reflects 
the dominant role of piezolelectric rates due to flexural modes, 
particularly at small doublet energies. In Fig. \ref{fig7}, we show 
the piezoelectric rates due to the flexural modes. The various 
dashed and dotted lines show the separate contributions for each 
mode, and the continuous line is the sum of all. Note that only the 
lowest flexural mode has a dispersion relation such that $\hbar 
\omega \to 0$ when $q_{\|} \to 0$ so that it contributes to all 
doublet separations. Higher modes tend to a finite energy when 
$q_{\|} \to 0$ and therefore their contributions start at increasing  
non-zero threshold values of $E_{if}$. Note that at small $|E_{if}|$ 
the long dashed line corresponding to the lowest flexural mode is 
indistinguishable from the sum due to the negligible contribution of 
the remaining modes. 

At the beginning of this section we have already explained the strong 
increase in this rate when $E_{if} \to 0$.  As the doublet separation 
increases the total rate decreases, with oscillations that only 
appear smooth due to use of a logarithmic plot, but which have 
significant features: at around $30 \,\mu$eV one sees a sharp jump 
due to the threshold of the third flexural mode and a pronounced dip 
is seen at $60 \,\mu$eV due to the oscillating contribution of the 
fifth mode. 
For the same flexural modes, the rates at small doublet energies are 
roughly eight orders of magnitude larger than those of the 
deformation potential interaction. Only at the threshold $\sim 0.1$ 
meV does the deformation rate reach a value comparable to the 
piezoelectric rate. The deformation rates might perhaps be more relevant 
in dots defined on a less piezoelectric semiconductor, such as InAs. 


{}Figs. \ref{fig9} and \ref{fig10} show the piezoelectric rates due 
to shear and dilatational modes. They are of similar magnitude to 
each other, and only beyond $50 \,\mu$eV do they become comparable to 
the flexural ones. Note also that they become very small at the 
lowest doublet separations. The deformation rates due to the 
dilatational modes are always one to two orders of magnitude smaller 
than the corresponding piezoelectric rates. 

\begin{figure}                   
\begin{center}
\leavevmode
\epsfxsize=8cm
\epsffile{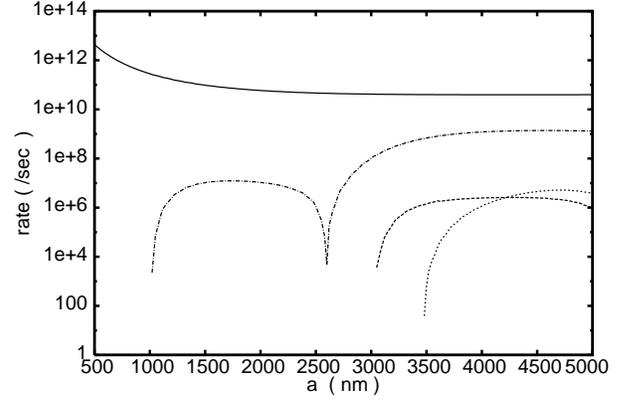}
\end{center}
\caption{ Piezoelectric rates due to flexural modes, for fixed $w_r = 
379.9 $ nm  ( $E_{if} = 6.1 \mu$eV ) , as a function of slab 
thickness $a$.  Continuous line: contribution of the lowest mode. 
Dashed lines :  other modes.} 
 \label{fig12}
\end{figure}

\begin{figure}                   
\begin{center}
\leavevmode
\epsfxsize=8cm
\epsffile{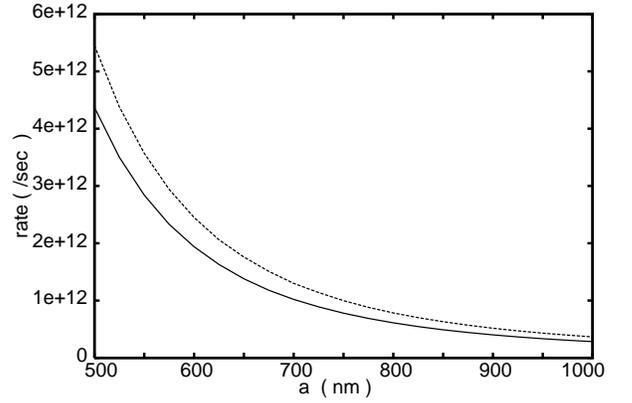}
\end{center}
\caption{ Piezoelectric rates for the lowest flexural mode, for fixed 
$w_r = 379.9 $ nm  ($E_{if} = 6.1 \mu$eV), as a function of slab 
thickness $a$;  ${\bar z}_0 = 70 $ nm (continuous line), 
${\bar z}_0 = 35 $ nm (dashed) .} 
 \label{fig13}
\end{figure}

\begin{figure}                   
\begin{center}
\leavevmode
\epsfxsize=8cm
\epsffile{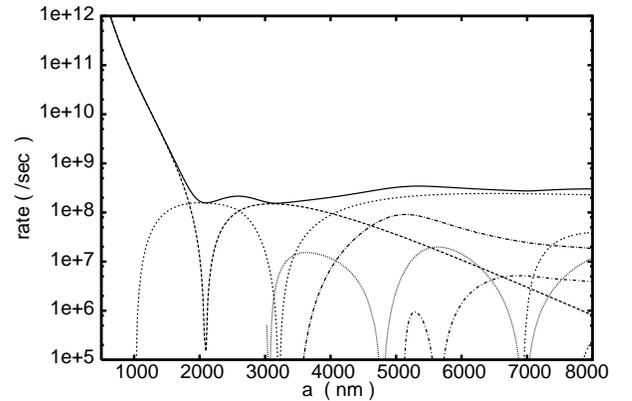}
\end{center}
\caption{ Piezoelectric rates due to the flexural modes, for 
$w_r = 379.9 $ nm  ($E_{if} = 6.1 \mu$eV) and ${\bar z}_0 = 
a/2- 180 $ nm, against $a$. 
Dashed lines:  individual modes; continuous line, total. } 
 \label{fig14}
\end{figure}

The dependence of the rates on slab thickness is shown in Figs. 
\ref{fig12} to \ref{fig14}. 
In Fig. \ref{fig12} we show piezoelectric rates due to 
flexural modes, for a fixed doublet splitting energy that corresponds 
to $w_r = 379.9 $ nm ($E_{if} \simeq 6.1 \,\mu$eV.) The contribution 
from the lowest flexural mode (continous line) dominates 
the others (dashed and dotted lines) over the whole range of 
thicknesses considered. Interestingly, increasing $a$ from 500 to 
2000 nm reduces the total rate by two orders of magnitude, showing 
the importance of finite slab thickness in determining rates.  

Fig. \ref{fig13} examines the dependence of that reduction on the 
2DEG depth. Bringing ${\bar z}_0$ closer to the surface increases the 
rates further, by about $25 \%$ in going from a depth of $70$ nm to 
$35 $ nm . Finally, Fig. \ref{fig14}  shows the difference 
in rates between  a ``surface'' and a ``bulk'' 2DEG: the location of 
the 2DEG has been set at ${\bar z}_0 = a/2 - 180$ nm and $a$ has 
been varied. The configuration, $a= 500$ nm, which sets ${\bar z}_0 
= 70 $ nm, was our standard configuration in all previous graphs. Now, as 
$a$ is increased, the 2DEG stays near the center of the slab, and its 
distance to the surface increases. One can see in Fig. \ref{fig14} 
two distinct regimes: ``surface'': for small  slab thicknesses the 
total rate due to all flexural modes decreases exponentially by as 
much as five orders of magnitude ; ii) ``bulk'' beyond $a \simeq 2000$ 
nm the rate stabilizes in order of magnitude, but oscillations of 
up to a factor of 2 still persist. The contributions to the rate due 
to shear and dilatational modes (not shown) are totally negligible at 
small $a$ and show a similar stabilization for $a$ above 2000 nm, 
where they are still one order of magnitude smaller than the flexural 
ones. Although experiments with a 2DEG at such depths are 
unrealistic, these results highlight the danger of 
estimating the rates for a double dot system without considering 
the influence of surface boundary conditions on the phonon spectrum. 

\section{Conclusions}
We have studied the main features of the tunneling rates between a 
doublet of states in a well-isolated two dot system defined on a slab 
of finite thickness. At low temperatures, $T = 100 $ mK, and for a 
piezoelectric material (GaAs),  our results confirm the dominance of 
the piezoelectric over the deformation rates. At small doublet 
separations we find that the lowest piezoelectric flexural mode gives 
the dominant contribution. Keeping the 2DEG at a fixed depth from the 
surface and increasing the slab thickness we find that the rates vary 
as much as two orders of magnitude, increasing as the slab thickness 
is decreased. Therefore in thin slabs the effective electron to 
acoustic phonon coupling is greatly enhanced. Complementarily, as a 
way to stress the relevance of boundary conditions on the phonon 
spectrum, when we keep the 2DEG near the middle of the slab and 
increase its thickness, we find that the ``bulk'' configuration leads 
to substantially smaller rates than the ``surface'' one. These 
findings should be useful in the design of future experiments on double 
dot systems supported on finite slabs.

\acknowledgements

Useful discussions with C.G. Smith are gratefully acknowledged. 
We are grateful to NSERCanada for Discovery Grant RGPIN-3198 (DWLS) 
and to DGES-Spain for continued support through grants  
BFM2001-3710 and FIS2004-03156 (JM). Guo Hong Yun is a visiting 
scholar at McMaster on research leave from IMU.


\end{document}